# Nodule detection and generation on chest X-rays: NODE21 Challenge

Ecem Sogancioglu, Bram van Ginneken, Finn Behrendt, Marcel Bengs, Alexander Schlaefer, Miron Radu, Di Xu, Ke Sheng, Fabien Scalzo, Eric Marcus, Samuele Papa, Jonas Teuwen, Ernst Th. Scholten, Steven Schalekamp, Nils Hendrix, Colin Jacobs, Ward Hendrix, Clara I Sánchez, Keelin Murphy

*Abstract*—**Pulmonary nodules may be an early manifestation of lung cancer, the leading cause of cancer-related deaths among both men and women. Numerous studies have established that deep learning methods can yield high-performance levels in the detection of lung nodules in chest X-rays. However, the lack of gold-standard public datasets slows down the progression of the research and prevents benchmarking of methods for this task. To address this, we organized a public research challenge, NODE21, aimed at the detection and generation of lung nodules in chest X-rays. While the detection track assesses state-of-the-art nodule detection systems, the generation track determines the utility of nodule generation algorithms to augment training data and hence improve the performance of the detection systems. This paper summarizes the results of the NODE21 challenge and performs extensive additional experiments to examine the impact of the synthetically generated nodule training images on the detection algorithm performance.**

## I. Introduction

Lung nodules may be an early manifestation of lung cancer, the biggest cancer killer among both women and men [1]. Because symptoms occur in late-stage disease and are aspecific, most lung cancers are diagnosed when the disease is already metastasized. However, the mortality rate varies significantly depending on the stage of the cancer when it was detected. While the 5-year survival rate of localized lung cancer is 59.0%, it is only 5.8% when the disease has metastasized [2]. This statistic highlights the crucial role of early detection of lung cancer in reducing mortality rates.

While CT scans are preferred over chest X-rays for lung cancer screening [3], [4], the inclusion criteria for CT screening programs are typically strict and a considerable number of patients who develop lung cancer in their lifetime might not be eligible for such screening programs. In contrast, Chest X-rays (CXR), being the most common imaging study acquired, play a crucial role for the detection of early lung nodules through routine clinical practice. Pulmonary nodules are frequently encountered as incidental findings in patients undergoing routine examination or CXR imaging for issues unrelated to lung cancer.

Superimposition of anatomical structures makes it challenging, however, to detect early lung nodules from a CXR as seen in Figure 1. In fact, several studies [5], [6] show that radiologist sensitivity for detecting nodules can vary from 36% to 84% on various datasets. Other work [7], [8] shows that 19%–26% of lung cancers visible on chest radiographs were, in fact, missed at their first readings.

Considering its high clinical relevance and potential impact, nodule localization has been one of the most widely studied topics on automated CXR analysis for decades [9]. This trend has changed in the last few years, however, with the release of publicly available CXR datasets (Chest X-ray14, CheXpert, MIMIC-CXR) [10]–[12] of which many publications made use [13]. The annotations of these datasets were obtained using natural language processing (NLP) techniques on radiology reports, and image-level labels are generated with more than 10 different abnormalities including lung nodules. The volume of publications inspired by these datasets demonstrates their value to the research community, particularly as large-scale training sets. For development of clinically applicable algorithms, however, evaluation must be extremely rigorous and evaluation dataset labels must be of a very high standard. Many of the works using these public datasets used the NLP-generated labels [14]–[18] for evaluation or, at best, radiologist assessment of CXR images [19]–[22]. The pitfalls of NLP labeling have been well documented [23], failing largely because the radiology report is not always a complete description of the entire image, but often refers only to a specific clinical question. Similarly named conditions such as pulmonary emphysema and subcutaneous emphysema are also known to be confused by such labelling systems. Radiological reading of CXR, while substantially better, also has limitations as a gold-standard; many nodules have a very subtle appearance on CXR and radiologist sensitivity and agreement is low in the absence of a CT or pathology-based gold standard.

For clinically relevant results a reference standard based on CT or on proven lung cancer is optimal, while algorithms should ideally pinpoint nodule locations to provide explainability and improve efficiency if acting as a second

This work was supported by the Dutch Technology Foundation STW, which formed the NWO Domain Applied and Engineering Sciences and partly funded by the Ministry of Economic Affairs (Perspectief programme P15-26 'DLMedIA: Deep Learning for Medical Image Analysis').

reader. Several studies have described work including such datasets used for evaluation of either a commercial or research algorithm for nodule localization [24]–[27]. However, since this data (or the evaluation platforms) remains inaccessible to the public, direct comparison with other nodule detection algorithms on the same dataset is not possible. The time, cost, and patient privacy issues associated with the collection of large datasets with strong reference standards limits the evaluation dataset possibilities for many researchers.

The data-hungry nature of modern deep-learning technologies means that researchers also require large training datasets. The labeling requirements in training data are generally less stringent than those in evaluation sets, however, NLP labels are generally insufficient for extracting useful cases of solitary pulmonary nodules, and radiological reading is certainly required to obtain training data for localization algorithms (those which identify the location of the nodule(s)). Given the time and costs associated with obtaining radiological labels and the difficulty with obtaining sufficient numbers of the most difficult cases (e.g. very subtle nodules or nodules concealed behind the diaphragm or the heart), there is strong motivation to investigate whether insertion of simulated nodules into CXR images could be a useful strategy to obtain training data. Such a dataset could be constructed to meet the needs of the user in terms of numbers of images, nodule sizes, locations, and conspicuity.

Motivated by these observations, we organized a public challenge, NODE21, which consists of two tracks: nodule detection and nodule generation. The aim of the NODE21 challenge is to improve the state of the art for the detection of solitary nodules on CXR. The nodule detection track assesses the performance of state-of-the-art nodule detection systems for CXR whereas the nodule generation track determines the utility of simulated nodule training data on the performance of nodule detection systems. Radiologist-annotated training data is made publicly available and private test sets have a CT-based reference standard. Algorithm evaluation is provided through the Grand-Challenge platform [28] and the challenge design ensures that the algorithms and code are publicly available and reproducible.

In this paper, we discuss the results of the detection and generation tracks of the NODE21 challenge. Additional extensive experiments are performed using various combinations of detection and generation algorithms to analyze the impact of the generated images on the detection performance and provide guidance on how best to incorporate simulated data as part of training data.

## II. Data

There are three datasets associated with NODE21: the training set, the experimental test set, and the final test set. The training dataset is made public, while both test sets are private and only visible to challenge organizers. Participants could evaluate their method against the experimental test set multiple times during the preparation of their algorithm to ensure correct working and to allow method or parameter tuning. Submission to the final test set was allowed only once, at the end of the challenge, to ensure that participants could not optimize their method for that set.

The aim of the NODE21 challenge is to improve the state of the art for the detection of solitary nodules on CXR. With this in mind, data selection for the challenge deliberately excluded images with a predominantly abnormal pattern of consolidation or infiltrates, nodular structures, clusters of nodules or more than three visible nodules. Nodules outside the size range 6mm-30mm were also excluded [29].

All datasets were pre-processed using the publicly available OpenCXR library [30] image standardization process. This process first removes homogeneous border regions and then applies energy-based normalization of image intensity values to standardize image appearance [31] using a lung segmentation [32] as an intermediate step. The images were then cropped to the region of the lung fields and resized to 1024 x 1024 pixels preserving aspect ratio and using padding on the shorter side. The training set is provided with both the original images and the pre-processed versions available, however, participants were advised that their algorithm would be tested on images that had been pre-processed in this way. Previous work [31] has demonstrated that such preprocessing makes image analysis systems more robust to variation in test data from different X-ray equipment, for example.

The NODE21 training dataset was made available on the Zenodo data sharing platform [33], with a Creative Commons Attribution-NonCommercial-NoDerivatives 4.0 International License.

The following subsections describe the three datasets in more detail.

### *A. Training dataset*

This dataset consists of postero-anterior (PA) chest radiographs (both with and without nodules) with bounding boxes provided to identify the nodule locations. The original data selection was made from public datasets where we had explicit permission to redistribute or where the dataset license provided permits it. These public datasets are as follows:

- JSRT [34]
- PadChest [35]
- ChestX-ray14 [36]
- Open-I [37]

To select images likely to contain nodules, data from each of these sets was chosen to include PA images with a label indicating nodule and (where possible) other labels selected to exclude confounding abnormalities such as consolidation or infiltrates. For a detailed description of the filtering process please refer to the annotation page [38].

Since the JSRT labels were provided by radiological examination with CT as the reference standard and with nodule location information available we did not re-label or re-annotate any of these cases, however, five cases were excluded as the nodules were outside the desired size range. All other selected data was reviewed in a reader study on the Grand-Challenge platform [28] by a chest radiologist with over 30 years of experience (ETS). The radiologist was asked to identify visible nodules by drawing bounding boxes around them. Images, where no nodule could be seen, where the nodule was outside the desired size range, or where there were significant confounding abnormalities, were excluded.

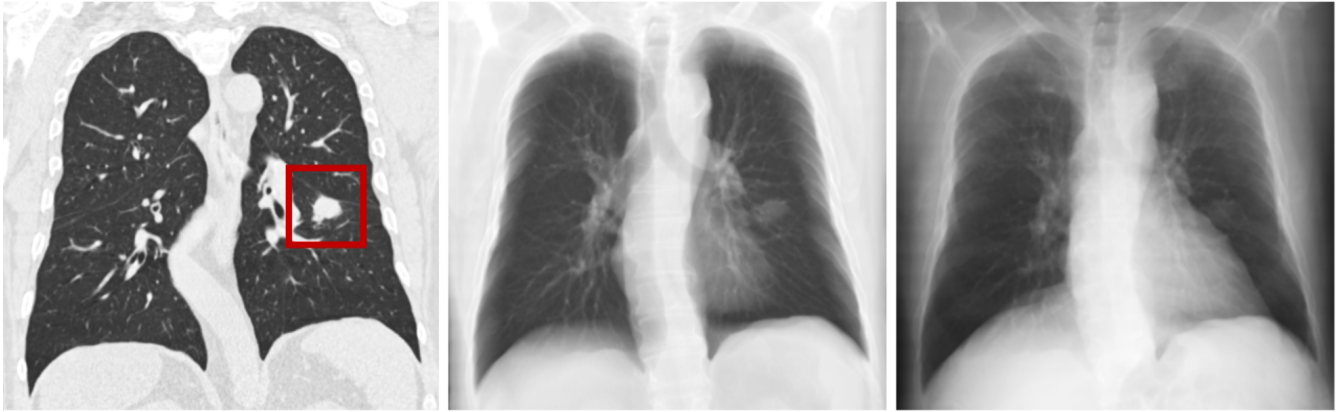

Fig. 1: Left: a coronal CT slice from a patient with a 2 cm squamous cell lung cancer (red box). On CT, this tumor is impossible to miss. Middle: this is the average of CT slices covering 10 cm of tissue centered on the slice on the left. The nodule is still clearly visible. Right: this is the average of all CT slices and similar to what a chest radiograph of this patient would look like. Surprisingly, even though this is a large nodule, not obscured by major organs like the heart, it is still only faintly visible due to all the other structures superimposed on it. This makes nodule detection from chest radiographs challenging.

In addition to the nodule images, a selection of normal images was also included for training. These images were chosen from the same four public datasets using the label of non-nodule for JSRT and of 'normal' or 'no finding' for the other datasets (please refer to [38] for a detailed explanation of filtering). Since the PadChest and ChestX-ray14 datasets had very large numbers of images matching the applied filters (28,688 and 22,452 respectively) a random selection of 1500 normal images was chosen from each in the initial selection. All selected images were then reviewed in a 3-step process as follows: 1) A member of the NODE21 team briefly reviewed each image and rejected any with obvious nodules or abnormalities. 2) The baseline nodule detection system (see section III-A for further details of this system) was run on the remaining images to identify suspicious regions, which were again reviewed by a team member. 3) Any case where the team member was uncertain whether a nodule was present was presented to a radiologist (ETS) for review and rejected if an abnormality was present.

Following this selection and review process, 1134 images (containing 1476 nodules) and 3748 non-nodule (normal) images were obtained. The numbers per dataset pre- and post-review are provided in Table I.

Participants were provided with the original CXR images (and identifiers to link them back to the original public sets) as well as with the OpenCXR standardized versions of the images.

For the generation track, we randomly sample 1000 images from the non-nodule images in the training dataset. Nodule bounding boxes were generated using our nodule location generator as shown in Figure 2. Up to 3 bounding boxes (1-3) were selected per image to be used to generate nodules inside. The images and the bounding boxes were provided to the participants at the test time where their submitted generation algorithm was expected to generate nodules inside the requested bounding boxes. Further, for the generation track, we provided a public set of NODE21 CT patches which participants were free to use as part of their generation algorithm. These are cropped 3D patches containing nodules from the public Luna16 CT dataset [39]. The patches were 50 x 50 x 50 mm, resampled to voxels of 1 x 1 x 1 mm. A total of 1186 nodule patches are provided together with associated nodule segmentations via the Zenodo data sharing platform [33].

TABLE I: Training data selection process. Nodule and non-nodule (normal) CXR images were selected from 4 public datasets: JSRT(J), PadChest(P), ChestX-ray14(C) and Open-I(O) and reviewed before inclusion in the challenge. Review steps are described in detail in the text. Figures indicate the number of images with the number of nodules in brackets.

| Source | J | P | C | O | Total(Nodules) |
|---|---|---|---|---|---|
| Nodule Data | | | | | |
| Initial Selection | 154 | 908 | 1586 | 82 | 2730 |
| After radiology review | 149 | 314 | 617 | 54 | 1134(1476) |
| Non-Nodule (Normal) Data | | | | | |
| Initial Selection | 93 | 1500 | 1500 | 1164 | 4257 |
| After 3-step review | 93 | 1366 | 1187 | 1102 | 3748 |

### B. Experimental test set

The images for the experimental test set were collected for this study is derived from standard clinical procedures at Radboud University Medical Center (RUMC) in Nijmegen, the Netherlands. For this set a reference standard of CT was required to confirm the presence of a nodule, so patients who had undergone both frontal CXR and a CT scan within a maximum of 60 days of each other were identified. These were further filtered to those patients whose corresponding CT report contained the word 'nodule'. A nodule detection system [40] was used to identify nodules on CT scans as a means of assisting the annotating radiologist (ETS). The radiologist was then provided with the 200 CXR images alongside the CT slices with detected nodules identified and asked to find the corresponding nodules on CXR and annotate them. The radiologist could also access the full CT scan if needed.

The annotator was asked to annotate solitary nodules, solid or subsolid, located in a region of otherwise normal appearance. Images with a predominant pattern of abnormal tissue, nodular structures, or clusters of nodules were excluded, as such very abnormal cases are not of interest. The cases where a nodule was only visible on CT but could not be seen on CXR even by estimating the approximate location were excluded from the study.

To select normal CXR images for inclusion in the experimental test set the Radboudumc CXR reports were searched for the text 'Normal image of heart and lungs' or the text 'Normal cardiopulmonary image' since these phrases had been observed to be used frequently to report a completely normal CXR image. From the results, a random selection of 120 PA CXR images from unique patients was made and provided to a chest radiologist for review. Images with any suspicion of nodule or other confounding abnormality were rejected.

A total of 281 CXR images were selected in this way, 166 of which contained (248) nodules and the remaining 115 were normal. The details are provided in Table II

### C. Final test set

The final test set was originally collected for a previous study [41]. It consists of 300 CXR images from 4 different hospitals in the Netherlands. A positive case is defined by the presence of a solitary pulmonary nodule visible on the PA image and confirmed by CT acquired within 3 months of the CXR. A negative case is one without nodules or other substantial pathology, confirmed by CT within 6 months of the CXR acquisition. The nodule locations were provided through the original study data and used for the NODE21 challenge. In addition, the findings of a total of 12 independent readers (6 radiologists and 6 radiology residents) are available, indicating whether or not they believe a nodule is visible on the CXR. These additional findings are used in this paper to provide an independent comparison between computer algorithms and human experts, but not as the reference standard for the challenge evaluation process. The final dataset excluded two images (which could not be obtained) from the original set in the paper and consists of 111 nodule-positive images (each with one nodule) and 187 non-nodule images as summarized in Table II.

### D. Additional Experiments Data

For additional experiments performed in this paper, in order to experiment with larger datasets, we utilized VinDr-CXR dataset [42]. VinDr-CXR is a publicly available dataset that contains 18k posterior-anterior view chest X-rays with both localization and classification labels for thoracic diseases. The images were labeled by a group of 17 experienced radiologists for the presence of 22 critical findings and 6 diagnoses. From this dataset, we selected 10606 images which were labeled with 'No finding', (meaning the selected CXR images are expected to contain no abnormalities). Nodule bounding boxes were generated using our nodule location generator as shown in Figure 2. Up to 3 bounding boxes (1-3) were selected per image to be used to generate nodules inside.

TABLE II: Experimental and Final test set statistics. Figures indicate the number of images with numbers of nodules in brackets. Before and After indicates the number of images before and after the radiology review, respectively. Details of the selection and review process are provided in the text.

|  |  | Nodule Data | Non-nodule Data |
|---|---|---|---|
| Experimental Test Set | Before | 200 | 120 |
|  | After | 166 (248) | 115 |
| Final Test Set | After | 111 (111) | 187 |

## III. CHALLENGE SETUP

The NODE21 challenge was hosted on the Grand-Challenge platform [28], which has hosted over 330 medical image analysis challenges since 2007. The challenge website is publicly accessible online at https://node21.grand-challenge.org/ ([29]), providing access to all information and functionality, including data, evaluation, and leaderboards. On Grand-Challenge, interested parties could register and find a general overview of the challenge including the deadlines, a description of the datasets, the evaluation metrics, and the preprocessing code. Through the website, the participants could submit their algorithms and access a forum to post questions or comments.

One aim of the NODE21 challenge was for competing algorithms to be fully reproducible and publicly available. To this end, only algorithm submissions in the form of a docker container were accepted. Docker containers encapsulate the software, allowing it to run uniformly and consistently on any system that supports Docker which plays a significant role in bridging the reproducibility gap in scientific research, reinforcing the credibility of the study, and fostering an environment conducive to collaborative scientific exploration and advancement. Once submitted, the container would automatically run on the private test set and generate results for the leaderboard. The submitted solutions were required to be linked to a public GitHub repository with a version tag and an Apache 2.0 or MIT license. The submitted algorithms are thus open-source and publicly available and can be tested out by interested users on the Grand-Challenge platform. Further, the GitHub repository details are visible for users interested in obtaining the code at https://github.com/node21challenge.

The NODE21 challenge was divided into two tracks, a detection track, where participants submitted algorithms for detecting nodules in CXR, and a generation track, where participants submitted algorithms for generating realistic nodules on normal CXR images. Interested parties could enter either or both tracks.

The challenge was open from October 19, 2021, when the training dataset was released, until January 25, 2022. Participants were allowed to submit their methods for evaluation on the experimental test set starting from December 2nd, to test their model performance as well as to make sure their docker submission worked as expected. Repeated submissions for evaluation on the experimental test set were permitted. From January 10th to 25th, participants were able to submit their final best algorithm to the final test set where only a single submission per participant was allowed. Submissions (to either phase) were not permitted after January 25, 2022.

### A. Baseline models

For each of the detection and generation tracks, a baseline model was provided with code available at https://github.com/node21challenge. This provided a benchmark performance for each track as well as template code for participants to demonstrate how to build working docker containers for submission to Grand-Challenge. The baseline methods are described in more detail in section IV-A

### B. Detection track

The detection track participants were required to submit an algorithm that reads a chest X-ray as input, and returns a list of bounding boxes for identified pulmonary nodules, with a likelihood score associated with each one.

*1) Evaluation Metrics:* For each algorithm submitted to the detection track, the following metrics were calculated: Area under the receiver operating curve (AUC) and sensitivities at average false positive (FP) rates of 0.125, 0.25, and 0.50 nodules per image.

To calculate the AUC, an image score was assigned to each chest X-ray by choosing the maximum bounding-box probability among detected nodules in that image. If there was no nodule prediction for an image, the image score was set at 0. These image scores were thresholded to obtain the receiver operating curve and, hence, the AUC.

To obtain sensitivities at different FP rates, free-response operating curve (FROC) analysis was used. If more than one predicted bounding box overlapped a reference bounding box with intersection-over-union (IOU) > 0.2 then only the prediction bounding box with the maximum probability among them was retained. Any prediction bounding box was then considered as a true positive if it overlapped with a reference standard bounding box at IOU > 0.2, otherwise, it was considered a false positive. Using the numbers of true and false-positives, we then calculated the average sensitivity at 3 predefined false positive rates: 1/8, 1/4, and 1/2 FPs per image. For cases where the FROC curve did not extend to the specified false positive rate the highest sensitivity value from the curve was used.

The final metric used to rank participants on the leaderboard was calculated as follows:

$$rank\_metric = (0.75 * AUC) + (0.25 * S) \quad (1)$$

where $S$ is the sensitivity at 0.25 FP per image. This gives a heavier weighting to the algorithm's ability to identify images containing nodules (which is the most clinically important task) but also considers its ability to correctly pinpoint the nodule locations.

### C. Generation track

The generation track participants were required to submit algorithms that take a frontal chest X-ray and one or more bounding box locations as inputs and return the same X-ray image with synthetically generated nodules inserted at the requested locations.

The locations of the nodules to be generated were predetermined by the challenge organizers. In order to select plausible locations on the input chest X-rays, deep learning based lung and heart segmentation algorithms [32] were run. The resulting segmentation maps were used to select the region where nodules could potentially appear, including the entire lung segmentation and the heart segmentation to the lowest detected point of each lung (see Figure 2). In order to include the lung regions obscured by the heart and diaphragm, for each lung the most upper point of the heart segmentation, the lowest detected point of the lung, and the leftmost point of the right lung (rightmost for the left lung) were used, which creates rectangular squared like region at the bottom of the lung areas. Up to three square bounding box locations (random from 1 to 3 nodules per image) with random sizes (7-17mm) were selected from this region and their locations and sizes were provided with the image for the generation algorithms to be trained with. We made sure that the nodule boxes fit inside the boundaries.

*1) Evaluation Metrics:* Generation track algorithms were evaluated by training a detection system with the generated images, including synthetic nodules, and evaluating the resulting nodule detection system as described in section III-B.1. This evaluation metric is based on the principle that a high-quality generation system should create images that can improve the performance of a detection system when included as training data.

For the evaluation of the generation algorithms, 1000 chest X-rays that are free of nodules were randomly selected from the NODE21 training dataset, and bounding box locations where nodules should be generated were pre-determined. This set of images and the locations were kept private and only visible to challenge organizers.

Once a generation algorithm was submitted, it was run on this dataset to output 1000 chest X-rays with generated nodules. The resulting generated images were used to train our baseline nodule detection system. This trained nodule detection system was then evaluated on the appropriate test set (experimental or final, depending on which phase the participant submitted to). The same evaluation metrics as used in the detection track (see section III-B.1 were calculated and detection performance was equated with generation performance for leaderboard ranking.

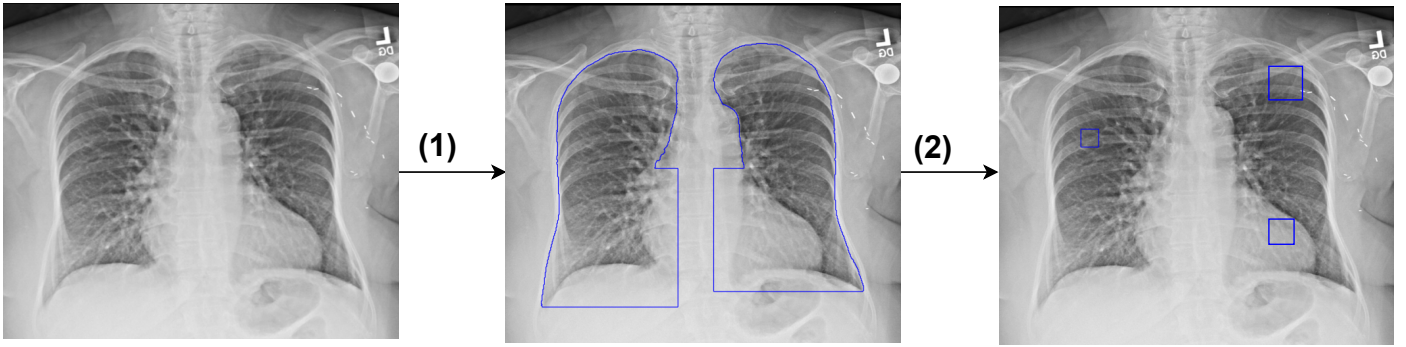

Fig. 2: Process used to identify locations where nodules should be generated. Step 1 applies heart and lung segmentation on a given CXR image and indicates the boundaries including heart and lungs to below the diaphragm. Step 2 receives the segmented CXR image and randomly places 1-3 square boxes (in the size range of 7-17mm) in the bounded regions.

## IV. CHALLENGE SUBMISSIONS

In total, 302 participants from various countries joined the challenge before the submission deadline. There were over 230 submissions to the experimental test set from both tracks combined. In the final test phase, 10 teams from 7 countries (6 teams for the detection track and 4 teams for the generation track) submitted a solution.

For inclusion in this paper, the best ranking methods from each track are selected for analysis and further experimentation. From the detection track, we include the top three performing methods as well as the baseline. Most of the generation track methods had a poor performance compared with the baseline and only one additional method (the top ranking method) was selected, along with the baseline, based on its methodology and performance.

### A. Baseline Methods

*a) DB (Baseline Detection Algorithm):* This model is the open-source baseline detection algorithm, which was provided by the challenge organizers before submissions were opened. It is based on a Faster R-CNN architecture [43] which uses ResNet50 [44] as the backbone. The model was trained on the OpenCXR-preprocessed version of the NODE21 training dataset.

In order to tackle the data imbalance issue, images with nodules were oversampled until the number of negative images was reached. The model was trained for 30 epochs, and early stopping was used in case of no improvement in the validation set performance for 5 consecutive epochs.

*b) GB (Baseline Generation Algorithm):* This model is the open-source generation algorithm that was provided by the challenge organizers before submissions were opened. The method requires 3D nodule templates segmented from CT scans. The algorithm is based on a simple cut and paste principle [45], [46], where nodules are generated from 3D nodule templates from CT scans and superimposed into a chest X-ray at the requested location. For each bounding box, a randomly selected nodule, which was cropped from a CT scan, was resampled so that it fit into the size of the given bounding box. As a next step, the resampled nodule was superimposed inside the bounding box, and the Poisson image blending technique [47] was applied to reduce local discrepancies around the corner regions.

This model used 3D nodule templates which were cropped from LUNA16 dataset and this dataset was also provided to the NODE21 participants together with the training dataset as described on Section II-A.

### B. Detection Track Top Submissions

In this section, we describe the top three detection solutions submitted. Methods D1, D2, D3 denote rank 1, rank 2 and rank 3 algorithms, respectively. Further details regarding the training strategies of these methods can be found online in the NODE21 challenge page.

*a) D1:* This model was placed as the rank 1 algorithm in the final leaderboard [48]. The submitted algorithm was an ensemble of 20 different models based on Faster RCNN [43], RetinaNet [49], YOLOv5 [50] and EfficientDet-D2 [51] architectures. Each model was trained using 5-fold cross-validation; Yolov5 were trained with a resolution of 640 x 640 and 1024 x 1024. The final ensemble used 5 models from each fold from Faster R-CNN, RetinaNet, Yolov5 (640 x 640 resolution), 4 models from Yolov5 (1024 x 1024 resolution), and 1 model from EfficientDet-D2 and the 20 model predictions were ensembled using weighted box fusion [52].

All the models were trained using the OpenCXR-preprocessed version of the NODE21 training dataset and no additional preprocessing steps were performed.

The Faster R-CNN and RetinaNet implementations utilized a pretrained ResNet-50 model as a backbone network. All the models except RetinaNet leveraged transfer learning, and pretrained the models on VinDr-CXR dataset [42]. All the model parameters were kept trainable during training.

In order to tackle data imbalance, the participants generated artificial nodules on 1000 randomly selected healthy images from the training dataset by using the GB.

For all the models except YOLOv5, various data augmentation schemes were applied such as cropping and padding, horizontal flipping, random rotation, blurring, and cutout augmentation. For the Yolov5 model, the original augmentation strategies were used, and test time data augmentation was applied.

*b) D2:* This model was ranked in 2nd place on the final leaderboard. The submitted algorithm is an ensemble of 33 YOLOv5 models, which were trained using 33 folds where each fold contains 85% of the nodules from the training dataset. The predictions of the 33 models were merged using a non-maximum suppression method.

Data balancing was tackled by undersampling the number of negative images in the training dataset. The model was trained from scratch using OpenCXR preprocessed version of the NODE21 training dataset. No further preprocessing steps were applied.

*c) D3:* This model was ranked 3rd on the final leaderboard. The algorithm is an ensemble of six models based on MaskRCNN and RetinaNet architectures with ResNet50 backbone. Three models per architecture were trained where each model was trained using different thresholding values for normalizing the dataset. The thresholding on the pixel intensities was performed based on predefined upper and lower quantile values, which was then followed by uniform normalization. Predictions of the six models were merged using non-maximum suppression.

All the models were trained using transfer learning. They were first pretrained with projected CT scans on the DeepLesion dataset [53], and then were further fine-tuned on the Luna16 dataset [39] to have better weight initialization. The resulting models were then trained on NODE21 training dataset where all the layers in the networks were kept trainable. Data imbalance was tackled by oversampling the positive cases for the training of Mask-RCNN model.

### C. Generation Track Top Submissions

The generation track aims to assess whether the state-of-the-art generation algorithms can improve the performance of the detection systems. The algorithms should take a frontal chest radiograph and one or more bounding box locations as input and produce an image with generated nodules at the requested locations.

*a) G1:* This model was placed as rank 1 algorithm in the final leaderboard. The nodule generation task was tackled by generative inpainting, where a network learns to inpaint the mask region in a given patch.

This model uses a generative adversarial networks (GANs), which consists of generator and discriminator networks specialized for inpainting. It is based on a recently proposed CR-Fill architecture [54], where the generator network receives a masked patch along with the actual mask and gradually produces the inpainted region. The generator has two components, the coarse network and the refinement network, where the predictions produced by coarse network are refined in the next step by the refinement network. Several losses were calculated to train the network; L1 loss was calculated from both the coarse and refined inpainted patches, adversarial loss and structural similarity index measure (SSIM) were calculated from the refined image patch. It also used contextual reconstruction loss from the feature maps produced by the refinement network, which aims to select useful patches from the image to fill in the missing region.

In order to increase the number of nodule cases to train the network, CheXpert [12] and MIMIC datasets [11] were utilized. Since these two datasets do not have location annotations, the baseline detection network was run on them, and predictions with confidence higher than 0.7 were selected. This procedure resulted in 7000 nodule images from CheXpert, 6500 nodule images from MIMIC. Since NODE21 dataset contains higher quality nodule annotations, the images from NODE21 were oversampled ten times during training. All the images were preprocessed using the OpenCXR library and no additional preprocessing step was performed. Horizontal flipping was used as augmentation during training.

## V. EXPERIMENTS

In addition to presenting the challenge evaluation metrics for each detection and generation method, in this work, we also evaluate an ensemble model of the best solutions in each track. The four detection track algorithms (D1-D3 and DB) were ensembled using the weighted box fusion method [52]. For the generation track, we combine all the generated images from both methods (G1 and GB) and assess the impact of this simulated nodule data in training. The performance of these experiments was evaluated in the same way as in the challenge. For the ensembled detection method, AUC score, and sensitivity at various false positive rate (0.5, 0.25, 0.125) were computed. In the generation track experiment, the combination of generated images produced by G1 and GB methods were used to train a baseline detection method. The resulting nodule detection system was then evaluated using the same detection evaluation metrics described above.

For this publication, additional experiments were performed to systematically assess the impact of the generated nodule images for building nodule detection systems. All models were trained from scratch without using any external data to make sure none of the images that were used in our test dataset. The experiments were designed to determine the impact of the dataset size, and the type of the detection and generation methods. In these additional experiments, we have utilized the large VinDr-CXR dataset [42] to generate nodules using G1 and GB methods aiding in assessing how dataset size affects performance. 10606 images which were labeled with 'No finding', were selected and up to 3 nodule bounding boxes (1-3) were generated using our nodule location generator(Figure 2, see Section II-D for details). Both G1 and GB generation methods were run on the images to synthesize nodules within the requested bounding boxes. These 10606 images are used for the experiments described in the remainder of this section to assess the impact of various factors when building a nodule detection system.

### A. Impact of the generation methods

These experiments aim to compare the performance of two nodule generation methods, namely G1 and GB, to create data for training nodule detection systems. To evaluate the utility of simulated nodules generated by these methods the baseline detection method, DB, was trained with their generated images only. Using a fixed detection method architecture allows us to investigate only the impact of the generation methods.

TABLE III: Network architecture and training details of NODE21 detection solutions. WBF= weighted box fusion, NMS= non max suppression

| Method | Architecture | Pretrained | Ensemble Size | Ensemble method | Input Resolution | Batch Size | Epochs | Data imbalance |
|---|---|---|---|---|---|---|---|---|
| D1 | YOLO FRCNN RetNet | VinDr-CXR | 20 | WBF | 1024 or 640 | 8 or 16 | 20-60 | simulated data |
| D2 | YOLO | ImageNet | 33 | NMS | 1024 | 8 | 60 | undersampling |
| D3 | MaskRCNN RetNet | DeepLesion Luna16 | 6 | NMS | 1024 | 8 | 40 | oversampling |
| DB | FRCNN | COCO | 1 | no | 1024 | 4 | 30 | oversampling |

In these experiments, the DB detection algorithm was trained solely with the generated CXR images (no real nodules) obtained using G1 or GB. For each generation model, we trained the detector firstly using all available images (10606 from VinDr-CXR), and further with various smaller dataset sizes set at 10%, 20%, 50% and 75% of the full simulated dataset. Finally, we used an ensemble approach and combined the images from both methods (G1+GB, resulting in 21212 images), and trained the detection model again using the full dataset and the specified subsets of 10%, 20%, 50%, and 75%.

The resulting detection models from each experiment were evaluated on the final test set, and AUC score and sensitivity at various false positive rate (0.5, 0.25, 0.125) were computed to measure the performance of the corresponding generation model or ensemble.

### B. Impact of the real dataset size

In these experiments, we consider the importance of the availability of real CXR nodule images and investigate the added value of generated nodule images for boosting model performance.

The NODE21 training dataset (4882 images, 1134 with nodules) is used as the source of real (not generated) nodule images, and 20k images with generated nodules are used (10606 VinDr-CXR images with nodules generated by each of 2 methods). The nodule images in the real dataset were oversampled until they reached the same size as the non-nodule images to have a balanced dataset. For each experiment where we combine real and generated data, we make sure that data was balanced as well by oversampling the real dataset size until it reaches the same size as the generated data. We investigate the impact of these generated images on the detection model when the number of real NODE21 images is varied.

The baseline detection method, DB, was trained using 10%, 20%, 50%, 75%, and 100% of the real dataset respectively. Next, each of these training datasets was combined with all the available generated images and DB was re-trained with each of these combination datasets. To prevent data imbalance where the number of real images was too small compared to the number of generated images, we oversample the real dataset until it reaches the generated dataset size during training.

The resulting detection methods from each experiment were evaluated on the final test set using AUC, and sensitivity at various false positive rates (0.5, 0.25, 0.125).

### C. Impact of combining detection and generation methods

In our final experiments, we analyze the impact of the generated data from different generators (G1 and GB) on each of the different detection methods described in this paper, namely D1, D2, D3, and DB. Each detection model was first trained with the real dataset (the NODE21 training dataset) to create a benchmark performance measure. Next, the detection method was trained by boosting the NODE21 training dataset with generated images, which were generated either by G1 or GB methods or a combination of both.

It is important to note that during the challenge the submitted algorithms were allowed to use external data sources and computational size or time was not limited for training. However, for these additional experiments, in order to compare the impact of the detection method (and not other factors such as external data or computational source), all three detection methods, D1, D2, and D3, were adapted to fit with the computational resource requirements. All models were trained from scratch without using any external data to make sure none of the images were used in our test dataset. The batch-size of the methods was decreased to be able to train each model with 12GB memory, and no other external data was used for training. For this reason, the benchmark performance measures are not identical to those achieved during the challenge. The specific modifications to each method were as follows: D1: The batch-size was reduced from 16 to 8 and the training set was limited to the provided NODE21 training data. D2: The batch-size was reduced from 16 to 8 D3: The method was trained with random weight initialization instead of pre-trained using external data-sources.

For all the experiments, we trained the corresponding detection model three times and the model with the best validation set performance was selected as the final model for evaluation. This was done in order to reduce the impact of the randomization process during training which arises from specific GPU computations.

The final model performance was evaluated on the final test set using AUC score, and sensitivity at various false positive rates (0.5, 0.25, 0.125).

TABLE IV: The performance of the nodule detection algorithms on the final test set of 298 images (111 with nodules). The ensemble model was obtained using weighted box fusion [52] on the four individual model predictions. AUC=Area under ROC curve, 'Outperforms' indicates the names of methods that have significantly lower AUC. S(0.5) indicates the algorithm sensitivity at an average of 0.5 false positives per image.

| Training Dataset Source : NODE21 training set | | | | |
|---|---|---|---|---|
| Training Dataset Size : 3748 images per generator | | | | |
| Test Dataset Source : NODE21 Final Test Set | | | | |
| Test Dataset Size : 187 (non-nodule) and 111 (nodule) images | | | | |
| Method | AUC | S (0.5) | S (0.25) | S (0.125) | Outperforms |
| D1 | 0.868 | 0.800 | 0.750 | 0.603 | DB |
| D2 | 0.862 | 0.771 | 0.723 | 0.600 | DB |
| D3 | 0.833 | 0.761 | 0.704 | 0.590 | |
| DB | 0.816 | 0.714 | 0.635 | 0.504 | |
| Ensemble | 0.877 | 0.819 | 0.754 | 0.619 | D3, DB |

TABLE V: The performance of the generation algorithms on the final test set. Each generator, G1, and GB methods were run on 1000 (non-nodule) CXR images from the training data, and the resulting data was used to train the baseline Faster R-CNN model. G1+GB denotes the experiments where images generated from both methods are combined (results in 2000 generated images). AUC=Area under ROC curve. 'Outperforms' indicates the names of methods that have significantly lower AUC. S(0.5) indicates the algorithm sensitivity at an average of 0.5 false positives per image.

| Training Dataset Source : NODE21 training set | | | | |
|---|---|---|---|---|
| Training Dataset Size : 1000 (generated-nodule) images per generator | | | | |
| Test Dataset Source : NODE21 Final Test Set | | | | |
| Test Dataset Size : 187 (non-nodule) and 111 (nodule) images | | | | |
| Nodule Generator | AUC | S (0.5) | S (0.25) | S (0.125) | Outperforms |
| G1 | 0.746 | 0.524 | 0.362 | 0.27 | |
| GB | 0.722 | 0.505 | 0.352 | 0.324 | |
| G1+GB | 0.783 | 0.591 | 0.51 | 0.463 | G1, GB |

## D. Statistical Methods

To compare AUC scores achieved by different methods, DeLong's test is used [55] with statistical significance set at $p<0.05$.

## VI. RESULTS

### A. Challenge Detection track

The performance of the detection algorithms on the final test data is provided in Table IV. As seen in the table, the D1 and D2 methods achieved a similar level of performance with an AUC of 0.868 and 0.862 and sensitivity of 0.800 and 0.771 at 0.5 FP rate, respectively. D3 method also achieved a high level of performance with an AUC score of 0.833 and sensitivity of 0.761 at 0.5 FP rate, however, only D1 and D2 algorithms showed significantly higher performance than DB ($p<0.05$).

The ensembled model which was created from the four detection track algorithms (D1, D2, D3 and DB) using weighted box fusion [52] achieved a significantly higher performance than D3 and DB methods ($p<0.05$) with an AUC of 0.877 and a sensitivity of 0.819 at 0.5 FP rate (Table IV). Further, the performance of the detection algorithms was compared to 12 experienced radiologists on the final evaluation set. Figure 5 illustrates the performance of each detection algorithm, the ensemble detection model, and the performance of the 12 observers. The final ensemble method showed a better performance than 3 radiologists and achieved a similar level of performance to 8 radiologists, underperforming only one radiologist.

Figure 3 shows example nodule cases from the final test set. As illustrated by the figure, while more obvious nodules are detected by all the detection methods, D1 and D2 methods perform better than D3 and DB methods for detecting small subtle nodules (nodules behind the rib and clavicles as in the example). Further, all the detection methods can miss very small or subtle nodules and nodules that appear in the region of vessels.

### B. Challenge Generation track

The generation algorithms were evaluated by running the baseline detection model with training data composed of CXRs with nodules generated by the submitted generation algorithm.

The baseline detection model trained with 1000 images generated by G1 and GB methods respectively achieved similar levels of performance with AUCs of 0.746 and 0.722 and a sensitivity of 0.524 and 0.505 at 0.5 FP rate, when evaluated on the final test data. These results are provided in Table V. The impact of combining images from both generation methods, G1, and GB, was evaluated by creating a total of 2000 nodule images (1000 images from each generation method). The final detection method (Faster R-CNN model) trained with these combined images achieved a significantly higher performance ($p<0.05$) than those trained using data from individual generation methods, with an AUC of 0.783 on the final test data (Table V.

In a qualitative evaluation, we visually inspected 200 randomly selected generated images from each method. Some examples of the generated images from both methods can be seen in Figure 4. While the G1 method consistently produced more visually realistic nodules compared to the GB method, the diversity of the nodule appearance (e.g. shape) was more limited. The GB method, on the other hand, tended to produce very bright nodules around the heart region and was more prone to producing extremely subtle nodules.

### C. Experiment Results

*1) Impact of the generation method:* The first set of experimental results where the baseline detection system was trained using solely generated data can be seen in Table VI. As seen in the table, the baseline detection method, DB, achieved an AUC of 0.701 and 0.722 using data from G1 and GB methods, respectively, and using only 10% of the generated data (1060 images). Considering networks trained using data from just one generator (G1 or GB), the AUC does not increase significantly when the dataset size is increased gradually from 10% all the way to 100% and remains in similar ranges ($p > 0.05$). Networks trained using G1 data alone also did not have a significantly different performance compared to those trained using GB data alone, regardless of dataset size.

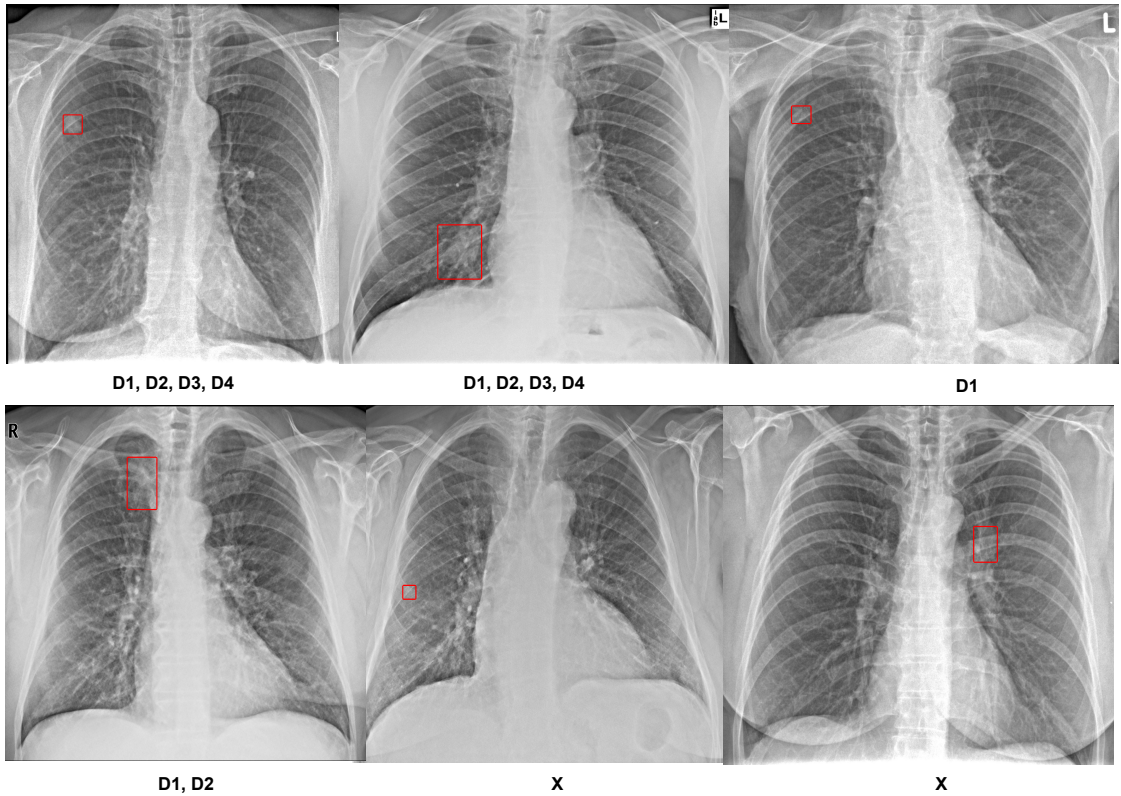

Fig. 3: Example nodule cases in the final test set. Detection methods that detected the corresponding nodule in the image are displayed below each image. X denotes that the nodule was missed by all the detection methods.

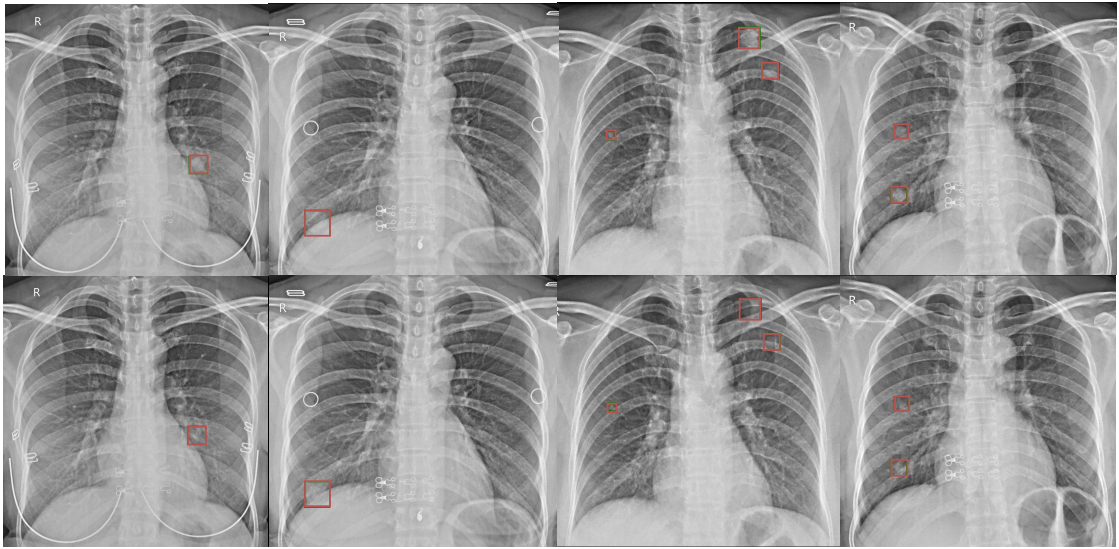

Fig. 4: Examples of generated nodules using G1 and GB methods. The top row indicates nodules generated by G1, and the bottom row shows generated nodules using GB.

It is, however, noteworthy that the performance consistently improved when G1 and GB generated datasets were combined for training, regardless of the dataset size. The highest performance levels (0.778-0.798) were achieved when the model was trained with the combination of G1+GB images, and the model trained with just 10% of this data (n=2,121) has an AUC that is not statistically different from that of the model trained with any other percentage, or with all data from both models (n=21,212).

*2) Impact of the real dataset size:* In Table VII, the results of varying the size of the real dataset available for training the

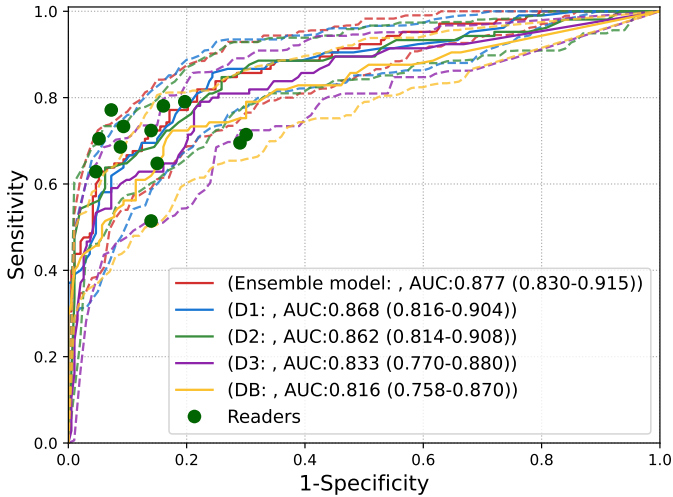

Fig. 5: ROC curve for the D1, D2, D3, and DB methods, their respective ensemble performance (denoted as ensemble model), and the 12 observers on the final test set of 298 images (111 with nodules). The ensemble model was obtained using weighted box fusion [52] on the four individual model predictions. AUC=Area under ROC curve.

DB detection model are shown. When training with real data only the detection performance improves consistently as the size of the dataset is increased.

Adding generated images into the training data results in performance improvements when only part of the real dataset is available. All the experiments except when the full real dataset was used showed significant improvement when generated images (G1+GB) were added ($p<0.05$).

The best performance was achieved when DB was trained using all NODE21 training data (real data) and all generated data which resulted in an AUC score of 0.844 and sensitivity of 0.762 at a 0.5 FP rate.

*3) Impact of combining detection and generation methods:* In this set of experiments, each of the detection algorithms (D1, D2, D3, and DB) was trained first using only real data (NODE21 training dataset), and then using a combination of real data with simulated nodule images generated by G1, G2 or both. The experiment results are displayed in Table VIII.

The AUC values achieved by models with generated data added to the NODE21 training data are generally not significantly better than those achieved by the model using only NODE21 (real) training data. The only exception to this is for D3, where adding all generated images (G1+GB) to the training set yielded a significant performance improvement($p<0.05$), increasing the AUC score from 0.766 to 0.812.

## VII. DISCUSSION

In this paper, we analyze the results of the two-track NODE21 challenge which was organized to collectively develop nodule detection and generation algorithms on chest X-rays. NODE21 was one of the first challenges which included only fully reproducible and open algorithm solutions. The best-performing algorithms from both tracks were selected to be included in the paper, and additional experiments were performed to systematically analyze the impact of the generated images on the state-of-the-art detection model performance.

The detection track solutions achieve results comparable to the 12 radiological readers (comprising six radiologists and six residents). Based on the 95% confidence intervals of the ROC curves, the ensemble model (of D1, D2, D3, DB) is outperformed by only one reader and its performance surpasses three others. Only two readers exceed the performance of the top model, D1, while its performance surpasses that of three others, as indicated by the 95% confidence intervals. The commercial nodule detection solution tested in 2014 on the same dataset [41] achieved a sensitivity of 81% at 1.9FP per image, while D1 alone can obtain 80% sensitivity at 0.5FP per image. This demonstrates the advances in AI in the last decade, which allow researchers to quickly develop systems with radiologist-level performance. Comparison with recent work using different test sets is generally difficult since the criteria for data selection and annotation vary widely. In 2021, a commercial system (AI Rad Companion Chest X-Ray algorithm (Siemens Healthineers AG)) achieved an AUC of 0.82 on a dataset that included CT-visualised nodules considered challenging to detect on CXR, similar to our test dataset [26]. This suggests that the detection networks in this challenge (AUC=0.816-0.868) perform in the range of the current state-of-the-art technologies.

Interestingly, the top three solutions (D1-D3) all utilized an ensemble of numerous state-of-the-art models (ranging from 6 to 33 models), implying the advantages of this strategy in achieving superior performance. These findings are consistent with prior studies [13], [56], and also align with the majority of the top-10 public challenge submissions such as CheXpert [12], SIIM-ACR [57], and RSNA-Pneumonia [58], all of which have utilized ensemble methodologies. Pinpointing the reasons for performance differences between the detection solutions (D1-D4) is a complex task, as there could be various factors at play. These may range from the quantity of models used in an ensemble, the architecture of the models themselves, to the strategies deployed during training. While the number of models in an ensemble appears to suggest an improvement in performance (Table III), this hypothesis requires additional validation through future research.

The generation track attracted a smaller number of participants and a scarcity of innovative methods. Most entries closely mirrored the provided baseline algorithm, with only one competitor (G1) employing a unique GAN method and demonstrating robust performance. This could suggest that the task of generating nodules was viewed as more challenging than that of nodule detection, a sentiment that aligns with previous research [13] where detection on CXR has been explored more comprehensively than the generation task.

Notably, our research demonstrates that combining nodules from different generators is considerably more beneficial than using a larger quantity from a single generator (as shown in Table VI). This underlines the importance of diverse generation techniques in real-world applications. The experiments, where the DB model was trained with a range of sizes of generated images (from G1, GB, or a combination of both), revealed that a model trained with only 2121 images from a combination of G1 and GB significantly outperforms one that is trained with a notably larger set of 10606 images from G1 alone. No model trained with data from a single generator could outperform a model trained on combined data, regardless of dataset size. One theory to explain this phenomenon is that when the data is generated using only a single method (G1 or GB), the detection method might learn features consistently produced by the generator, such as sharp nodule borders for example, which are not always present in real nodule data. We hypothesize that generating data with different methods can help the detection algorithm to focus less on these generator-specific features, and more on the important nodule characteristics. This theory is consistent with previous work on natural images [46] which showed that combining generated images with different blending techniques performed better than using a single blending technique for an object localization task.

TABLE VI: Impact of the generation methods: The baseline detection model, DB, is used; G1 and GB denote 10606 images with nodules generated by G1 and GB models, respectively. Each experiment was run three times, and the model with the best performance on the validation set was selected. AUC=Area under ROC curve. 'Outperforms' indicates which methods which have significantly lower AUC. (Methods that were never significantly worse than others are not listed in these additional columns). * indicates that the method with significantly worse performance was trained with a dataset that was larger or the same size. S(0.5) indicates the algorithm sensitivity at an average of 0.5 false positives per image.

| Training Dataset Source : VinDr-CXR dataset | | | | | | | | | | | | | | |
|---|---|---|---|---|---|---|---|---|---|---|---|---|---|---|
| Training Dataset Size : 10606 (generated-nodule) images per generator | | | | | | | | | | | | | | |
| Test Dataset Source : NODE21 Final Test Set | | | | | | | | | | | | | | |
| Test Dataset Size : 187 (non-nodule) and 111 (nodule) images | | | | | | | | | | | | | | |
| Dataset | AUC | S (0.5) | S (0.25) | S (0.125) | Nb of images | Outperforms | | | | | | | | |
| | | | | | | 10% | | 20% | | 50% | | 75% | | 100% |
| | | | | | | G1 | GB | G1 | GB | G1 | GB | G1 | GB | G1 | GB |
| 10% G1 | 0.701 | 0.392 | 0.349 | 0.295 | 1060 | | | | | | | | | | |
| 10% GB | 0.722 | 0.504 | 0.400 | 0.307 | 1060 | | | | | | | | | | |
| 10% G1 + GB | 0.778 | 0.581 | 0.495 | 0.381 | 2121 | ✓ | ✓ | ✓* | | ✓* | | ✓* | | ✓* | |
| 20% G1 | 0.708 | 0.438 | 0.371 | 0.324 | 2121 | | | | | | | | | | |
| 20% GB | 0.734 | 0.476 | 0.419 | 0.359 | 2121 | | | | | | | | | | |
| 20% G1 + GB | 0.797 | 0.590 | 0.472 | 0.381 | 4242 | ✓ | ✓ | ✓ | ✓ | ✓* | | ✓* | | ✓* | |
| 50% G1 | 0.716 | 0.466 | 0.383 | 0.324 | 5303 | | | | | | | | | | |
| 50% GB | 0.743 | 0.505 | 0.429 | 0.343 | 5303 | | | | | | | | | | |
| 50% G1 + GB | 0.797 | 0.648 | 0.505 | 0.416 | 10606 | ✓ | ✓ | ✓ | ✓ | ✓ | | ✓ | ✓ | ✓* | |
| 75% G1 | 0.700 | 0.414 | 0.390 | 0.343 | 7954 | | | | | | | | | | |
| 75% GB | 0.743 | 0.571 | 0.457 | 0.354 | 7954 | | | | | | | | | | |
| 75% G1 + GB | 0.798 | 0.619 | 0.552 | 0.475 | 15909 | ✓ | ✓ | ✓ | ✓ | ✓ | ✓ | ✓ | ✓ | ✓ | ✓ |
| G1 | 0.709 | 0.405 | 0.381 | 0.362 | 10606 | | | | | | | | | | |
| GB | 0.745 | 0.545 | 0.466 | 0.352 | 10606 | | | | | | | | | | |
| G1+GB | 0.782 | 0.638 | 0.524 | 0.438 | 21212 | ✓ | ✓ | ✓ | | ✓ | | ✓ | | ✓ | |

TABLE VII: Impact of the real dataset size: The baseline detection model, DB, is used; G1 and GB denote 10606 images with nodules generated by G1 and GB models, respectively. Each experiment was run three times, and the model with the best performance on the validation set was selected. AUC=Area under ROC curve. S(0.5) indicates the algorithm sensitivity at an average of 0.5 false positives per image. * indicates that the AUC is significantly improved compared to the previous row (without generated training data)

| Training Dataset Source : VinDr-CXR and NODE21 training dataset | | | | | |
|---|---|---|---|---|---|
| Training Dataset Size : 10606 (generated-nodule), 4882 NODE21 images | | | | | |
| Test Dataset Source : NODE21 Final Test Set | | | | | |
| Test Dataset Size : 187 (non-nodule) and 111 (nodule) images | | | | | |
| Dataset | AUC | S (0.5) | S (0.25) | S (0.125) | Nb of images |
| 10% Real | 0.742 | 0.409 | 0.314 | 0.276 | 488 |
| 10% Real+ G1 + GB | 0.802* | 0.668 | 0.584 | 0.499 | 21700 |
| 20% Real | 0.774 | 0.571 | 0.438 | 0.350 | 976 |
| 20% Real+ G1 + GB | 0.830* | 0.704 | 0.629 | 0.556 | 22188 |
| 50% Real | 0.789 | 0.676 | 0.584 | 0.504 | 2441 |
| 50% Real + G1 + GB | 0.848* | 0.795 | 0.635 | 0.523 | 23653 |
| 75% Real | 0.797 | 0.685 | 0.590 | 0.512 | 3661 |
| 75% Real + G1 + GB | 0.849* | 0.779 | 0.693 | 0.584 | 24873 |
| Real | 0.816 | 0.714 | 0.635 | 0.504 | 4882 |
| Real + G1 + GB | 0.844 | 0.762 | 0.638 | 0.514 | 26094 |

TABLE VIII: Impact of the generated images on the performance of detection methods. Real = 4882 NODE21 images, G1 and GB are 10606 images with simulated nodules generated by G1 and GB methods, respectively. AUC=Area under ROC curve. S(0.5) indicates the algorithm sensitivity at an average of 0.5 false positives per image. T denotes training time in days. *indicates the AUC values found to be significantly different (comparisons were only made within each detection method)

| Training Dataset Source : VinDr-CXR and Node21 dataset | | | | | |
|---|---|---|---|---|---|
| Training Dataset Size : 10606 (generated-nodule), 4882 NODE21 | | | | | |
| Test Dataset Source : NODE21 Final Test Set | | | | | |
| Test Dataset Size : 187 (non-nodule) and 111 (nodule) images | | | | | |
| Method | AUC | S (0.5) | S (0.25) | S (0.125) | T (days) |
| **D1** | | | | | |
| Real | 0.845 | 0.781 | 0.707 | 0.594 | 2 |
| Real+G1 | 0.838 | 0.759 | 0.674 | 0.543 | 5 |
| Real+GB | 0.844 | 0.792 | 0.713 | 0.584 | 5 |
| Real+G1+GB | 0.852 | 0.784 | 0.733 | 0.600 | 6 |
| **D2** | | | | | |
| Real | 0.846 | 0.782 | 0.704 | 0.619 | 1 |
| Real+G1 | 0.849 | 0.784 | 0.712 | 0.639 | 2 |
| Real+GB | 0.851 | 0.779 | 0.711 | 0.623 | 2 |
| Real+G1+GB | 0.858 | 0.762 | 0.705 | 0.609 | 2 |
| **D3** | | | | | |
| Real | 0.766* | 0.614 | 0.449 | 0.352 | 1 |
| Real+G1 | 0.778 | 0.639 | 0.599 | 0.502 | 1 |
| Real+GB | 0.789 | 0.645 | 0.601 | 0.521 | 1 |
| Real+G1+GB | 0.812* | 0.695 | 0.614 | 0.533 | 1 |
| **DB** | | | | | |
| Real | 0.816 | 0.714 | 0.635 | 0.504 | 4h |
| Real+G1 | 0.832 | 0.724 | 0.648 | 0.505 | 6h |
| Real+GB | 0.827 | 0.719 | 0.644 | 0.500 | 6h |
| Real+G1+GB | 0.844 | 0.762 | 0.638 | 0.514 | 6h |

Another insight from our results is that increasing generated dataset size does not significantly increase the performance of the detection model (for either single generators or combined). Using only 10% of the available data produces a comparable result to using 100% for all datasets (G1, GB, G1+GB). This indicates that the generated nodules are useful but lack diversity, hence producing larger numbers of them does not aid performance. However, it is notable that DB detection systems trained only with synthesized nodule images can lead to high-performance levels (with AUC close to 0.8 for those trained with combined datasets) when evaluated on real nodule images. This highlights the usefulness of the generation dataset.

In the next set of experiments (Table VII), the focus was placed on examining the enhancement provided by the inclusion of synthetically generated images when real data (chest X-rays with nodules) was accessible. For models using up to 75% of the real dataset, there was a statistically significant improvement in performance when the 21k synthetic nodule images were added to the dataset. When using the full real dataset for training the baseline model, the addition of the generated data did not result in a statistically improved model, even though the AUC value did rise from 0.816 to 0.844. Notably, however, this latter version of the DB model achieved comparable performance to that of D1 and D2 ($p > 0.05$) (in contrast to the original DB which had significantly worse performance compared to D1 and D2 (Table IV). This indicates that the addition of the generated data has elevated the performance of the simple Faster R-CNN model to be comparable to that of much more complex ensemble models.

The finding that generated data is most useful when real dataset size is limited is consistent with the additional findings in Table VIII. While adding G1 and GB datasets into training yielded a slight increase in AUC for all the detection methods, only D3 showed a statistically significant improvement when retrained with all generated nodule data added to the real training dataset. This suggests that for the other three methods, the real data contains sufficient data diversity for the task.

In conclusion, the results from the NODE21 challenge demonstrate that the utilization of generated nodule data can improve the effectiveness of detection methods for identifying nodules in CXRs under certain scenarios. Given the data and methodologies applied in this study, the enhancement was most prominent when the size of the real dataset was restricted and when data generated from two different generation methods was combined. These findings suggest that employing various generation methods, or possibly even differing method parameters or blending techniques, can increase diversity and likely offer more benefits compared to merely using the same generator to produce a larger volume of images. Future efforts should concentrate on enhancing the diversity of the generated images to potentially achieve greater advances in performance.

## VIII. Acknowledgment